\documentclass[useAMS,usenatbib,usegraphicx]{mn2e}

\title[XMM-Newton X-ray Observations of the Wolf-Rayet Binary System WR 147]
{XMM-Newton X-ray Observations of the Wolf-Rayet Binary System WR 147 }
\author[S.L. Skinner, S.A. Zhekov, M. G\"{u}del, and 
        W. Schmutz] 
{S.L. Skinner$^{1}$\thanks{E-mail:
skinners@casa.colorado.edu (SLS); szhekov@space.bas.bg (SAZ);
guedel@astro.phys.ethz.ch (MG); werner.schmutz@pmodwrc.ch (WS)},
S.A. Zhekov$^{2}$, M. G\"{u}del$^{3}$, and W. Schmutz$^{4}$ \\
$^{1}$CASA,  Univ. of Colorado, Boulder, CO 80309-0389 USA \\
$^{2}$Space Research Institute, Moskovska str. 6, Sofia-1000, Bulgaria \\
$^{3}$Paul Scherrer Institute, W\"{u}renlingen and Villigen,
     CH-5235 Switzerland \\
$^{4}$Physikalisch-Meteorologisches Observatorium Davos, Dorfstrasse 33, 
      CH-7260 Davos Dorf, Switzerland}
\begin{document}

\date{Accepted 2007 April 25. Received 2007 March 26; 
in original form 2007 March 26.}

\pagerange{\pageref{firstpage}--\pageref{lastpage}} \pubyear{2002}

\maketitle

\label{firstpage}

\begin{abstract}
We present results of a $\approx$20 ksec X-ray observation
of the Wolf-Rayet (WR) binary system WR 147 obtained with 
{\em XMM-Newton}. Previous studies have shown that this
system consists of a nitrogen-type WN8 star plus an OB
companion whose winds are interacting to produce a colliding
wind shock.  X-ray spectra from the pn and MOS detectors
confirm the high extinction reported from IR studies and
reveal hot plasma including the first detection
of the Fe K$\alpha$ line complex at 6.67 keV. Spectral fits
with a constant-temperature  plane-parallel shock model
give a shock temperature kT$_{shock}$ = 2.7 keV 
(T$_{shock}$ $\approx$ 31 MK), close to but slightly hotter
than the maximum temperature predicted for a colliding wind 
shock. Optically thin plasma models
suggest even higher temperatures, which are not yet ruled out.
The X-ray spectra are
harder than can be accounted for using 2D numerical 
colliding wind shock models based on nominal mass-loss parameters.
Possible explanations include: (i) underestimates of the 
terminal wind speeds or wind abundances, (ii) overly simplistic
colliding wind models, or (iii) the presence of other X-ray
emission mechanisms besides colliding wind shocks. Further
improvement of the numerical models to include potentially
important physics such as non-equilibrium ionization will
be needed to rigorously test the colliding wind interpretation.
\end{abstract}

\begin{keywords}
stars: individual (WR 147) -- stars: Wolf-Rayet -- X-rays:stars.
\end{keywords}

\section{Introduction}
The WR binary system WR 147 (AS 431) has been observed
extensively at optical, infrared, and radio wavelengths.
The southern  component WR 147S is a WN8 star (van der
Hucht 2001).
Near-infrared images clearly reveal a second source (WR 147N)
located $\approx$0.64$''$ north of the WN star which was
classified as a B0.5V star by Williams et al. (1997, 
hereafter W97). An earlier O8-O9 V-III spectral type
was proposed on the basis of {\em HST} observations
(Niemela et al. 1998). The E(B$-$V) values of WR 147N
and WR 147S are similar (Niemela et al. 1998), suggesting
that they are physically associated.

The  distance determined from 
near-IR photometry by  Churchwell et al. (1992,
hereafter C92) is 630 $\pm$70 pc, which we adopt
here for consistency with our previous {\em ASCA}
$+$ {\em VLA} study (Skinner et al. 1999, hereafter
S99). At this distance the projected separation of
0.64$''$ equates to 403 $\pm$ 45 AU. The physical 
separation depends on the orbital inclination angle 
which is uncertain. An early study based on analysis
of {\em VLA} radio images gave 
$i$ = 45$^{\circ}$ $\pm$ 15$^{\circ}$ 
(Contreras \& Rodriguez 1999) but more recent 
work suggests a lower inclination consistent with  
$i$ $\approx$ 30$^{\circ}$ (Dougherty et al. 2003;
Contreras, Montes, \& Wilkin 2004).

Near-simultaneous {\em VLA} observations at five 
different frequencies (S99) clearly show that WR 147S
has a strong wind. It is a thermal radio  source and its
flux density S$_{\nu}$ increases  at high frequencies 
according to S$_{\nu}$ $\propto$ $\nu^{+0.62}$,
as expected for  free-free wind emission. 
The radio-derived ionized mass-loss rate of the
WN8 star based on its 1.3 cm flux density and assuming 
a spherical homogeneous wind
(Wright \& Barlow 1975) is in the range
$\dot{M}$(WR) =  (1.9 - 4.0) $\times$ 10$^{-5}$
M$_{\odot}$ yr$^{-1}$ (S99). The  lower
rate is based on solar abundances and the higher
value assumes generic WN abundances.

The radio emission at lower frequencies is dominated
by a nonthermal component (C92; Contreras \& Rodr\'{i}guez
1996; Moran et al. 1989; S99). The IR and radio
images of W97 suggest that the nonthermal 
emission peak lies about 0.$''$07 south of
the IR source WR 147N.  This offset is very small
but does raise the interesting possibility that
the nonthermal radio emission originates in a
colliding wind shock standing off the surface 
of the OB star. 

X-ray emission from WR 147 was detected 
by the {\em Einstein}
observatory (Caillault et al. 1985) and moderate
resolution CCD spectra were obtained with  
{\em ASCA} (S99). A high spatial resolution 
{\em Chandra} HRC-I image analyzed by 
Pittard et al. (2002) did not provide 
spectral information but did show that
the X-ray emission peak lies at or near the
northern source WR 147N. It thus appears
that the X-rays originate either in the colliding 
wind shock region near the OB star or in
the OB star itself, and both may contribute.

An analysis of the
{\em ASCA} spectra showed that the X-ray
emission is dominated by a relatively cool
plasma component at kT$_{1}$ $\approx$ 1 keV,
but a hotter component at kT$_{2}$ $>$ 2 keV
could not be ruled out (S99). It was shown
that a colliding wind shock could  account 
for the temperature of the cooler plasma in 
the WR 147 spectrum but the X-ray luminosity
predicted by the colliding wind shock model 
was a few times larger than observed.

We have obtained higher signal-to-noise observations
of WR 147 with {\em XMM-Newton} that provide improved
information on emission lines and X-ray temperature
structure. The {\em XMM-Newton} spectra clearly reveal 
a high-temperature (kT$_{2}$ $>$ 2 keV) component 
in addition to the cooler plasma detected by {\em ASCA}.
Here, we revisit the interpretation of the 
X-ray emission in light of the new data and summarize
results from shock models and discrete-temperature
optically thin plasma models.

\section[]{Observations}

The {\em XMM-Newton} observation  began on
2004 Nov. 4 at 22:54 UT and ended on Nov. 5 at 05:12 UT.
Pointing was centered on WR 147 (J203643.65$+$402107.3).
The  European Photon Imaging Camera (EPIC) provided 
CCD imaging spectroscopy from the pn camera 
(Str\"{u}der et al. 2001) and two nearly
identical MOS cameras (MOS1 and MOS2;
Turner et al. 2001). 
The medium optical blocking
filter was used. The EPIC cameras provide
energy coverage over E $\approx$ 0.2 - 15 keV with 
energy  resolution E/$\Delta$E $\approx$ 20 - 50.
The pn and MOS camera pixel sizes are 4.$''$1 and
1.$''$1 respectively. 
The MOS cameras provide the best on-axis angular 
resolution with FWHM $\approx$ 4.3$''$ at
1.5 keV.

Data were reduced using the {\em XMM-Newton}
Science Analysis System (SAS vers. 7.0). The SAS
pipeline processing scripts {\em epchain} and
{\em emchain} were executed in order to incorporate
the latest calibration files available as of January 2007.
Event files generated by these scripts were time-filtered
to remove $\approx$4 ksec of PN exposure and $\approx$3 ksec
of exposure per MOS in the middle of the observation
that were affected by high background radiation. 
After removing the high-background segments we obtained
16.95 ksec of usable pn exposure and 19.64 ksec of
exposure per MOS.

Spectra and light curves were extracted from  circular
regions of radius  R$_{e}$ = 25$''$ centered
on WR 147,  corresponding to $\approx$80\%
encircled energy at 1.5 keV. 
Background files were extracted from circular 
source-free regions near the source. 
The number of net counts in the background-subtracted
spectra were: 4436 (pn), 2085 (MOS1), and 2063 (MOS2).
Spectra were binned
to a minimum of 20 counts per bin for analysis.
The SAS tasks {\em rmfgen} and {\em arfgen} were used to
generate source-specific response matrix
files (RMFs) and auxiliary response files
(ARFs) for spectral analysis. The data were
analyzed using the {\em XANADU} software
package
\footnote{The {\em XANADU} X-ray analysis software package
is developed and maintained by NASA's High Energy
Astrophysics Science Archive Research Center (HEASARC). See
http://heasarc.gsfc.nasa.gov/docs/xanadu/xanadu.html
for further information.},
including {\em XSPEC} vers. 12.3.0.

\section{Observational Results}

We present here the main  observational results 
including images, light curves and spectra.

\subsection{X-ray Images}

WR 147 was clearly detected by all three EPIC
cameras. Figure 1 shows a broad-band (0.5 - 7 keV)
EPIC pn image of WR 147 as well as a narrow-band
(6.3 - 7 keV) pn image. The latter is dominated by
the FeK$\alpha$ emission-line complex at 6.67 keV, 
which is clearly seen in the spectra (Sec. 3.3).

The centroid position of WR 147 in the broad-band pn 
image is RA = 20$h$ 36$m$ 43.67$s$, Decl. =
$+$40$^{\circ}$ 21$'$ 07.1$''$ (J2000.0), which 
is offset by only 0.$''$3 from its {\em HST}
Guide Star Catalog (v2.2)  position
(J203643.65$+$402107.3).  We have compared
the centroid position of WR 147 in soft (0.5 - 3 keV)
and hard (2 - 7 keV) images and find no significant
offset. The offset of the X-ray centroid in the 
FeK$\alpha$ image (Fig. 1) from the Simbad 
position is only  0.$''$5, which
is less than  the raw (unbinned) pixel 
size.

\subsection{X-ray  Light Curve}

Figure 2 shows the broad-band (0.5 - 7 keV)
EPIC pn light curve of WR 147. The mean pn count 
rate is 0.25 $\pm$ 0.03 (1$\sigma$) counts s$^{-1}$. 
A fit of a constant count rate model
to the entire pn light curve (including the 4.1 ksec
high-background segment) binned at 400 s intervals
gives a probability of constant count rate
P$_{const}$ = 0.45 ($\chi^2$/dof = 51.5/51).
The same fit applied to the last 9.1 ksec of
pn data after the high-background subsided gives
P$_{const}$ = 0.54 ($\chi^2$/dof = 40.4/42). 
Thus, we find no evidence of significant 
variability down to 400 s time intervals.

\subsection{X-ray Spectra}

Figures 3 and 4 show the EPIC pn and MOS1 spectra
of WR 147. The spectra are heavily absorbed below 1 keV
as previously observed with {\em ASCA}.
This absorption is anticipated on the basis of 
the known high extinction toward WR 147 
(A$_{\rm V}$ =  11.5  mag, C92; 
A$_{\rm V}$ =  11.2  mag, Morris et al. 2000,
hereafter M00).

Strong line emission is detected. The most 
prominent lines along with their laboratory
energies and maximum line power temperatures are 
Si XIII (E$_{lab}$ = 1.865 keV, log T$_{max}$ = 7.0), 
S XV (E$_{lab}$ = 2.46 keV, log T$_{max}$ = 7.2), 
and the FeK$\alpha$ complex which includes
Fe XXV (E$_{lab}$ = 6.67 keV, log T$_{max}$ = 7.6). 
A faint feature that we classify as a possible 
line detection is  Mg XI (E$_{lab}$ = 1.35 keV, 
log T$_{max}$ = 6.8). Also, there is a hint of
Ar XVII (E$_{lab}$ = 3.13 keV, log T$_{max}$ = 7.3)
in the MOS spectra, but if present this feature 
is very weak. 

The FeK$\alpha$ line is a new detection and was not 
seen by {\em ASCA}. We have generated simulated
{\em ASCA} spectra based on the {\em XMM-Newton}
results which show that the inability to detect the 
FeK$\alpha$ line with  {\em ASCA}  was
likely a result of insufficient sensitivity. 
As Figure 1 shows, the FeK$\alpha$
emission clearly originates in the WR 147 
system and is not background related. The
presence of the FeK$\alpha$ feature leaves no doubt
that very high temperature plasma ($>$10$^{7.4}$ K)
is present in this binary system.

\section{Stellar Properties}

The assumed values of stellar properties relevant to 
our analysis are summarized below.

\subsection{Abundances}

We assume solar abundances for the OB star
and adopt the values of Anders \& Grevesse (1989). 
Abundances of the WN star are expected to be nonsolar
because of its advanced evolutionary state.
The X-ray spectra do not constrain the 
abundances of H, He, C, N, O and their values
were held fixed during spectral fits at values
appropriate for WR 147 (Table 1 notes). The 
H and He abundances are from M00 and 
give the number ratio He:H = 2.5:1.
The adopted  C, N, and O abundances are  
generic WN abundances  taken from 
van der Hucht et al. (1986; hereafter vdH86)
and reflect the strong overabundance of
nitrogen anticipated in WN stars.  

The abundances of  Ne, Mg, Si, S, Ar, Ca, and Fe 
do in general affect the X-ray fits and their
abundances were allowed to vary in the fitting 
procedure. The starting values for the abundances
of Mg, Si, S, Ar, and Fe were initialized at
the canonical WN abundances of vdH86 
and those of Ne and Ca were
initialized to the values determined for WR 147
by M00. No obvious X-ray emission lines are 
detected from  Ne or Ca and their abundances
are not well-constrained by the X-ray data.  

\subsection{Mass-loss  Parameters}
To maintain consistency
with earlier work, we have adopted the same mass-loss
parameters as in S99, which are:
$\dot{M}$(WR) = 4 $\times$ 10$^{-5}$ M$_{\odot}$ yr$^{-1}$,
$v_{\infty}$(WR) = 950 km s$^{-1}$,
$\dot{M}$(OB) = 6.6 $\times$ 10$^{-7}$ M$_{\odot}$ yr$^{-1}$,
$v_{\infty}$(OB) = 1600 km s$^{-1}$. The value
$v_{\infty}$(WR) = 950 km s$^{-1}$ is identical to a 
more recent measurement by M00. The above mass-loss
parameters, which we herafter refer to as the
{\em nominal} mass-loss parameters, give a ratio 
of wind ram pressures
P$_{WR}$/P$_{OB}$ $\approx$ 36, placing the 
wind interaction region at a position coincident
with that of the nonthermal radio source near
WR 147N. The wind momentum ratio for the above
mass-loss parameters is 
$\eta$ = 
[$\dot{M}$(OB)$v_{\infty}$(OB)]/[$\dot{M}$(WR)$v_{\infty}$(WR)] 
= 0.028, in agreement with that of Niemela et al. (1998). 

\subsection{Extinction}
The IR studies   of M00 and C92 give A$_{V}$ = 11.2 mag
toward WR 147. Our X-ray emission models include an absorption
component based on Morrison \& McCammon (1983)
cross-sections.  The X-ray models provide a best-fit
value for the neutral hydrogen column density
N$_{\rm H}$, which  was allowed to vary during the fits. 
We obtain an equivalent A$_{V}$ using the Gorenstein (1975)
conversion N$_{\rm H}$ = 2.22 $\times$ 10$^{21}$~A$_{V}$~cm$^{-2}$. 
The A$_{V}$ determined from X-ray spectra is in good 
agreement with IR values.

\section{Spectral Analyis}

We summarize here spectral modeling results that seek to
determine whether the observed X-ray emission is consistent
with that predicted for a colliding wind shock.

\subsection{Plane-Parallel Shock Models }
We have  fitted the pn and MOS spectra with  a  constant
temperature plane-parallel shock model $vpshock$
(Borkowski et al. 2001) in XSPEC using the latest
APED ionization fraction data (neivers 2.0). This
is a generic shock model in the sense that it does not
directly incorporate mass-loss or orbital data for
WR 147. However, the model does account for sophisticated
physics including non-equilibrium ionization (NEI) effects.
The $vpshock$ model assumes equal electron and ion 
temperatures. 

As Figure 5 shows, the $vpshock$ model does a remarkably 
good job of reproduding the EPIC spectra. The best-fit
parameters are given in Table 1. The inferred N$_{\rm H}$
gives A$_{V}$ = 10.0 [9.1 - 11.4; 90\% conf.] mag using
the Gorenstein (1975) conversion. This value is slightly
less than A$_{V}$ = 11.2   determined from IR studies but is consistent
with  IR results to within 90\% confidence limits.

The shock temperature kT = 2.7 [2.4 - 2.9; 90\% conf.] keV 
inferred from $vpshock$ is slightly larger than
expected from  CW estimates for WR 147.
The maximum shock temperature on the line of
centers for an adiabatic shock is 
kT$_{cw}$ = 1.95$\mu$$v_{1000}^2$ 
(Luo et al. 1990), where $\mu$ is the mean
particle weight (ions and electrons)
in the shocked plasma and 
$v_{1000}$ is the shock velocity in units
of 1000 km s$^{-1}$. For the WN abundances
adopted here,  $\mu$ = 1.16.
Thus, kT = 2.7 keV implies a pre-shock
wind velocity $v$ = 1093 [1030 - 1132] km s$^{-1}$.
This is only 15\% higher than the currently
accepted value for the WN star (950 km s$^{-1}$).
Given the difficulty in accurately determining
terminal wind speeds and element abundances, 
this agreement is quite good.

The best-fit $vpshock$ model slightly underestimates
the flux below 1 keV,  more noticeably in
the pn fit than in MOS (Fig. 5). This suggests that excess soft
emission could be present that is not accounted for
by the high-temperature shock, perhaps originating
in the  OB or WN stars. This motivates us
to consider a two-temperature model that also 
includes a cool emission component.

\subsection{Discrete Temperature Models}
It is possible that the observed X-ray spectrum
is the superposition of a cool stellar component
and hot colliding wind shock plasma.
A cool X-ray component (kT$_{1}$
$<$ 1 keV) may originate in the winds of the OB
or WN stars themselves via radiative instability
shocks (Lucy 1982; Lucy \& White 1980). These
early studies considered forward shocks in the wind
but later work suggests that stronger reverse shocks 
may also be present (Owocki, Castor, \& Rybicki 1988).
At the spatial resolution of {\em XMM-Newton}, any X-ray
emission arising from shocks in the winds of the
individual stars cannot be separated from colliding
wind X-ray emission.  

To test the idea that the observed X-ray emission is
the superposition of cool  stellar radiative wind shock
emission plus a hotter colliding wind component, 
we have fitted the EPIC spectra
with a discrete temperature optically thin plasma
$vapec$ model using two temperature components (2T $vapec$). 
The results of this model are shown in Figure 6 and 
summarized in Table 1. There is indeed some improvement
in the fit below 1 keV but overall the fit statistic
is nearly identical to that of the plane-parallel
shock model. Formally, we cannot distinguish between
the two based on goodness-of-fit. For comparison, 
we note that an isothermal optically thin plasma
model (1T $vapec$) does not produce an acceptable fit.

The absorption N$_{\rm H}$ determined from the 2T $vapec$
model is identical to that of $vpshock$ and implies
A$_{V}$ = 10.0 [9.5 - 10.9; 90\% conf.] mag. 
The  derived temperature for
the  cool component kT$_{1}$ = 0.76 keV  is 
realistic for radiative-driven wind shocks.

The most  noticable difference between the $vpshock$
model and 2T $vapec$ is that the latter gives a 
considerably higher temperature for the hot component
kT$_{2}$ = 3.6 [3.2 - 4.1; 90\% conf.] keV. If due to
colliding winds, this would imply a pre-shock wind velocity 
of at least 1260 km s$^{-1}$. This velocity is
33\% higher than the currently accepted terminal wind
speed of the WN star (950 km s$^{-1}$). As such, 
the 2T  $vapec$ model is more difficult to reconcile
with the colliding wind picture unless the terminal
wind speed of the WN star is considerably higher 
than currently believed or the mean particle weight
in the WR wind is well above the adopted value $\mu$ = 1.16.

\subsection{Numerical Colliding Wind Models}

The models discussed above do not explicitly account
for the colliding wind geometry of WR 147. The geometry
and temperature structure of a colliding wind shock 
as well as its intrinsic X-ray  luminosity 
depend on the  mass-loss
parameters of the WR and OB components as well as the
binary separation. Thus, a realistic model must take
these into account.

We have produced synthetic X-ray spectra of WR 147
based on 2D numerical hydrodynamic simulations that
directly incorporate mass-loss parameters and binary
separation. Our approach is similar to that outlined
in previous studies (S99,  Myasnikov \& Zhekov 1993,
Zhekov \& Skinner 2000).
However, some improvements have been made including
use of the latest atomic data in the APED
\footnote{http://cxc.harvard.edu/atomdb/} data base.
The numerical modeling work is ongoing and further
refinements are needed but the summary below does provide 
some insight that may be  valuable as a 
guide for future work.

This model assumes that {\em all} of the observed
X-ray emission originates in an adiabatic colliding wind
shock, which could be an oversimplification if the stars
themselves also contribute. The model assumes equal
electron-ion temperatures in the shocked WN and OB star
winds.  This is appropriate for the 
WN star (eq. [1] of Zhekov \& Skinner 2000) but 
electron-ion temperature differences could be important
in the shocked OB star wind. However,  our results show 
that the shocked OB wind contributes only a small fraction 
to the absorbed X-ray flux, amounting to $\approx$12\% in
the 0.5 - 10 keV range and $\approx$18\% in the 
4 - 10 keV range. The numerical simulation also 
assumes ionization equilibrium and  in this respect is
different than the plane-parallel shock model $vpshock$,
which  does account for non-equilibrium ionization (NEI).
Incorporation of NEI effects into the numerical hydrodynamic
model is computationally expensive but may eventually be
necessary to rigorously test the colliding wind model.
In our initial simulations we have assumed that the
physical separation is equal to the projected 
separation D = 403 AU and have used the mass-loss parameters
given in Section 4.2. But, we do  consider 
below  how deviations
from these values affect the simulated spectra. 

There are two differences between the existing colliding
wind models and the data that need to be reconciled before
the models can be considered satisfactory. First, the 
models predict an intrinsic X-ray luminosity that is 
$\approx$3.5 times larger than inferred from the spectral 
data. Second,
the models do not account for all of the hard-band flux
seen in the spectra in the range 4 - 7 keV (Fig. 7). 
That is, the observed spectrum is harder than our  models predict
using nominal mass-loss parameters.

The X-ray luminosity mismatch was also apparent in our 
analysis of the {\em ASCA} spectra (S99). This difference
could easily be accounted for by uncertainties in the 
distance, mass-loss parameters and orbital separation. The
colliding wind X-ray luminosity scales with mass-loss rate 
$\dot{M}$, wind speed $v$, and separation D
as L$_{x}$ $\propto$  $\dot{M}^2$$v^{-3.2}$D$^{-1}$
(Luo et al. 1990; Stevens et al. 1992). The mass
loss rates could be a factor of  $\sim$2 lower than
assumed  if the winds are clumped (M00),
which would account for the difference. But, a
factor of two reduction is not even  required
since the true binary separation will be larger
than the projected  separation 403 AU due to 
inclination effects. Using the inclination 
estimates cited in Section 1, $i$ $\approx$ 30$^{\circ}$
gives D = 465 AU and would require that the assumed  
mass-loss rates be scaled down by a factor of 
$\approx$1.7, while $i$ $\approx$ 45$^{\circ}$
implies D = 570 AU and gives a scale factor of
$\approx$1.6. For the above scale factors, one would obtain
$\dot{M}$(WR) = (2.3 - 2.5) $\times$ 10$^{-5}$ 
M$_{\odot}$ yr$^{-1}$.

The fact that the observed spectrum is harder in
the 4 - 7 keV range than our current models predict
may hold important clues to colliding wind physics
and stellar properties.
We are able to produce a harder synthetic X-ray
spectrum by assuming higher terminal wind speeds.
If the WR and OB star wind speeds
are increased by 30\% to 
$v_{\infty}$(WR) = 1235 km s$^{-1}$  and
$v_{\infty}$(OB) = 2080 km s$^{-1}$ then 
the fit in the 4 - 7 keV range is substantially
improved, as shown in Figure 7.  But the fit is
still not  statistically acceptable.
Other factors such as NEI effects may
thus be important and possible electron-ion temperature
differences would need to be accounted for at
higher wind speeds and shock temperatures. 
Implementation of NEI effects in the numerical
colliding wind models is beyond the scope of
this study and will be  addressed in future
work.

\subsection{Summary of Spectral Model Results}

The best spectral fit from the models considered 
above is obtained with a
plane-parallel constant-temperature shock 
model ($vpshock$). The inferred shock temperature 
kT = 2.7 keV (T $\approx$ 31 MK) is close to,
but slightly greater than, the maximum colliding
wind shock temperature predicted for nominal
mass-loss parameters and the adopted  abundances.
A 2T optically thin plasma model (2T $vapec$)
gives nearly as good a fit but converges to a hot-component 
temperature kT  =  3.6 keV (T $\approx$ 42 MK).
This higher temperature is more difficult to reconcile
with colliding wind models based on current wind speed
estimates, but is not ruled out. Preliminary colliding
wind models  based on numerical hydrodynamic simulations
do not give acceptable fits and further refinements
are needed to rigorously test the colliding wind
predictions.

\section{Discussion}

We discuss below the implications of the new X-ray
results and compare them with previous observations. 

\subsection{Long-Term X-ray Behavior of WR 147}

A comparison between the X-ray properties of WR 147 determined
from {\em ASCA} spectra obtained in 1995 (S99) and those 
derived from this {\em XMM-Newton} observation
nine years  later is informative.  Overall, the 
best-fit {\em XMM-Newton} models give results that are
in very  good agreement with lower signal-to-noise 
{\em ASCA} data.

The N$_{\rm H}$ values determined by {\em ASCA} and
{\em XMM-Newton} are identical. The fluxes determined
from  {\em XMM-Newton} depend somewhat on the model 
used and on which of the three EPIC spectra are fitted.
If all three EPIC spectra are fitted simultaneously
(Table 1) then the  absorbed X-ray flux is 6\% less than 
the {\em ASCA} value that was determined from a 
2T $mekal$ model and the EPIC unabsorbed flux is
33\% larger than {\em ASCA}. If only the higher
signal-to-noise EPIC pn
spectrum is fitted, then the unabsorbed fluxes 
between {\em XMM-Newton} and {\em ASCA} agree
to within 1\%. Based on these comparisons, we find
no compelling evidence that the X-ray flux of 
WR 147 has changed significantly in the nine years
since the  {\em ASCA} observation.

However, sampling of the X-ray emission of WR 147
in the time domain is sparse.
Further X-ray monitoring on shorter timescales
of $\sim$months to a year might be worthwhile to 
determine if short-term X-ray variability 
is present. Changes in the centimeter radio
flux density of WR 147S by  $\approx$25\% 
on a timescale of $\approx$1 year have been 
reported by Contreras \& Rodr\'{i}guez (1999)
using matched-beam {\em VLA} observations.
If these changes reflect a change in the 
mass-loss rate of the WN star then the 
X-ray luminosity might also be affected,
since as already noted (Sec. 5.3), the colliding
wind X-ray luminosity is dependent on the mass-loss
rate. Changes in orbital separation D can also 
affect L$_{x}$, but these changes will be negligible
on the timescales considered here because of the
wide binary orbit.

\subsection{X-ray Temperatures: How Hot is WR 147?}

The nominal wind parameters for the WN star (Sec. 4.2)
and mean particle weight $\mu$ = 1.16 give a predicted
maximum temperature for the shocked WN star wind
in a colliding wind shock
kT$_{cw}$ = 1.95$\mu$$v_{1000}^2$ = 2.04 keV
or T$_{cw}$ = 24 MK.  Colliding
wind theory predicts  the hottest plasma
will lie on the line-of-centers between the
two stars and cooler plasma off the 
line-of-centers (Luo et al.1990;
Stevens et al. 1992). 

Several lines of evidence suggest that 
the plasma is actually hotter than the
above calculation predicts. Both the 
$vpshock$ and 2T $vapec$ models converge
to higher temperatures than above, and the 
spectrum is harder than numerical 
colliding wind models can account for
using nominal wind parameters. Very
hot plasma is clearly present since the
Fe K$\alpha$ line emits maximum power
at T$_{max}$ $\sim$ 40 MK but the line 
can form over a range of temperatures.

If the plasma is in fact hotter than predicted
by colliding wind models, how can it be 
explained? One possible explanation is that 
the adopted wind speeds or abundances are too
low.  As we have shown, increasing the wind
speeds by 30\% significantly improves the
spectral fit in the 4 - 7 keV range (Fig. 7). 
Terminal wind speeds
are difficult to determine observationally
because for WR stars the wind continues to accelerate
beyond radii where spectroscopic wind-speed 
diagnostics are available (Schmutz 1997).  

It could also be that the numerical colliding
wind models are oversimplified and don't fully
account for important physics. The existing 
colliding wind models do not incorporate  NEI 
effects. However, the plane-parallel shock 
model $vpshock$ does and it provides a better 
fit to the high-energy part of the spectrum.
This could be a clue that NEI effects
are important. This can be the  case in
young shocks that have not had time to reach 
ionization equilibrium and the plasma is 
underionized, or in steady-state shocks
just downstream from the shock front.
NEI can affect the strength of forbidden lines in
He-like triplets and grating spectra may 
be needed to determine if NEI is in fact present.

Lastly, physical mechanisms other than colliding winds
may be at work.  Nonthermal processes such as inverse
Compton scattering could give rise to hard X-ray
(or $\gamma$-ray) emission, but there is no strong
justification for invoking nonthermal emission since
the X-ray emission at higher energies can be fitted
satisfactorily with a thermal model (Fig. 6). In
addition,  Reimer, Pohl, \& Reimer (2006) have argued
that nonthermal emission would likely be masked by
stronger thermal emission in WR 147 and difficult 
to detect.  

A more intriguing possibility is that hot thermal
plasma at temperatures of $\sim$30 - 40 MK due to 
magnetic processes may be present. Such high
temperatures are quite typical of   magnetic
reconnection processes in lower mass stars.
Hot plasma that may be of magnetic origin
has also been detected in some high mass stars 
such as the B0.2V  star $\tau$ Sco 
(Cohen et al. 2003). And, the detection of
nonthermal radio emission from WR 147N
is a signature of magnetic phenomena (S99).
The possibility that magnetic phenomena are
contributing to the high-temperature X-ray
emission detected in WR 147 thus remains open,
but is still  speculative.

\subsection{Multi-Temperature Plasma}

It is somewhat surprising that the constant
temperature $vpshock$ model provides such
a good fit of the spectrum. In the 
colliding wind picture, the plasma is
not expected to be constant-temperature
since hotter plasma lies along the
line-of-centers.
The presence of emission lines in the
spectra that form over a range of 
temperatures (Sec. 3.3) is a clear
indication that the X-ray emitting plasma 
is not isothermal.

It is also obvious that any soft emission 
(kT $<$ 1 keV) that may be present from 
radiative-instability shocks
in the winds of the OB or WN stars is not
accounted for in the constant-temperature
$vpshock$ model. This may explain why 
the $vpshock$ model, as well as the
colliding wind model, underestimate
the flux at lower energies near 1 keV 
(Figs. 5 and 7). The 2T $vapec$ model 
places most (82\%) of the
emission measure in cool plasma at
kT$_{1}$ = 0.8 keV (9 MK) and does
a somewhat better job accounting for
the emission below 1 keV, but some
low-level ($\approx$1$\sigma$) fit
residuals are still present.

\subsection{Implications of Clumped Winds}

There is now observational evidence suggesting
that the winds of massive WR and O stars are clumped
rather than homogeneous. This has been discussed
in detail for WR 147 by M00. The most important
implication of clumped winds for our X-ray
analysis is that the mass-loss rates could be
a few times lower than the nominal (unclumped)
values given in Section 4.2.  If the mass-loss
rates are lower then the predicted X-ray
luminosity from colliding winds is also lower since
L$_{x}$ $\propto$  $\dot{M}^2$$v^{-3.2}$D$^{-1}$.
However, the importance of this effect in wide
binaries such as WR 147 depends on the ability of
clumps to survive at large distances from the WR
star  and on their survivability within the wind
interaction region itself (Pittard et al. 2006).
As we have already noted (Sec. 5.3), the difference
in L$_{x}$ between that inferred from the X-ray
apectra and colliding wind predictions could be
accounted for by a lower mass loss rate. Some
X-ray evidence for lower mass-loss rates consistent with
clumped winds has also been
presented for the WC $+$ O binary $\gamma^2$ Velorum 
(Schild et al. 2004). But, we do emphasize 
that L$_{x}$ also scales inversely with  
the binary separation D, which is not well-known
for WR 147 because of uncertainties in the orbital inclination.
A more accurate determination of the inclination is
needed to determine the factor by which  the 
homogeneous wind mass-loss rates
(Sec. 4.2) should be scaled down to obtain agreement
between predicted and observed L$_{x}$.

Other factors besides clumping also enter into
the uncertainty in mass-loss rates. In particular,
the uncertain spectral type and luminosity class of
the OB companion (Sec. 1) give rise to uncertainties
in its  mass-loss rate. We have investigated
the sensitivity of our numerical colliding wind
models to changes in the assumed OB star mass-loss rate
by factors of up to $\pm$2 relative to the 
nominal values. Such variations have very little
effect on  the shape of the simulated 
spectrum or the $\chi^2$ fit statistic since
the shocked WR wind dominates the X-ray emission.

The theory of clumped winds predicts that stochastic
X-ray bursts should occur if a clump in the WR
wind collides with a clump in the O wind (Cherepashchuk 1990).
This offers, at least in principle, an indirect  means
of detecting wind clumps at X-ray wavelengths and 
quantifying their properties. However, in  practical
terms the detection of wind clump X-ray bursts 
would be a major observational challenge. Clump-clump
collisions should be rare (Cherepashchuk 1990) and
long-term time monitoring would be needed. Also, high
signal-to-noise X-ray light curves would be required to
discriminate between random noise and clump-induced
stochastic variability. If such a search were conducted,
WR 147 would not necessarily be an optimal target because
of its   wide binary separation.
At a projected binary separation of  403 AU and an assumed 
stellar radius R$_{\rm wr}$ $\approx$ 20 R$_{\odot}$ (M00)
the wind-wind collision zone occurs at more than
4000 stellar radii from the WR star. The clump 
number density at such large distances would be reduced
due to increasing volume and beyond some distance the wind
will start to look approximately homogeneous. Furthermore,
there are legitimate  questions as to whether clumps could 
even survive as coherent structures so far out in the WR wind.
Closely-spaced WR $+$ O or WR $+$ WR binaries 
would seem to be more promising candidates for 
a clump-induced X-ray variability search.

\section{Summary}

We have presented results of a sensitive
{\em XMM-Newton} X-ray observation of WR 147.
The main results are the following:

\begin{enumerate}

\item \noindent The X-ray spectrum of  WR 147 shows 
      strong absorption below $\approx$1 keV and clear
      evidence for high-temperature plasma, including 
      a first detection of the
      Fe K$\alpha$ emission line complex.  \\

\item \noindent The extinction  
      A$_{V}$ = 10.0 [9.1 - 11.4] mag determined from
      the X-ray absorption is in good agreement with 
      IR estimates. \\

\item \noindent  No significant X-ray variability was seen over
      the $\approx$20 ksec observation. The observed (absorbed) X-ray
      flux agrees to within a few percent with       
      that measured by {\em ASCA} nine years earlier. \\

\item  \noindent A constant-temperature plane-parallel shock model
      gives a shock temperature kT$_{shock}$ = 2.7 keV
      (31 MK) which is slightly higher than predicted  by colliding
      wind theory. An equally satisfactory  two-temperature 
      optically thin plasma model
      implies even higher temperatures, which are not yet ruled out. 
      Some cooler plasma (kT $<$ 1 keV) may also be present that
      most likely originates in either the WN or OB stars, but
      the X-ray spatial resolution is not sufficient to resolve the
      system into individual binary components. \\

\item \noindent The X-ray spectrum is harder than predicted by
      2D numerical colliding wind shock models using currently accepted
      wind parameters. This could reflect inaccurate wind speed
      or abundance estimates, simplifying assumptions in the 
      numerical models, or hard X-ray production via physical 
      processes other than colliding winds. Further refinements
      in the numerical hydrodynamic models are needed to 
      incorporate potentially important physics such as 
      non-equilibrium  ionization effects.

\end{enumerate}

\section*{Acknowledgments}

This research was supported by NASA grants NNG05GB48G and 
NNG05GE69G.
Work at PSI (M.G.) was supported by  Swiss National Science 
Foundation grant 20-66875.01. 
This work is based on observations
obtained with {\em XMM-Newton}, an ESA science mission with instruments
and contributions directly funded by ESA states and the USA (NASA).

\newpage
\clearpage

\begin{table*}
 \centering
 \begin{minipage}{140mm}
  \caption{XMM-Newton Spectral Fits for WR 147}
  \begin{tabular}{@{}lll@{}}
  \hline
Parameter                             &                  &      \\
                                      &                  &      \\
\hline
Spectrum                              & pn$+$mos         & pn$+$mos         \\
Model                                 & vpshock          & 2T vapec    \\
N$_{\rm H}$ (10$^{22}$ cm$^{-2}$)     & 2.2 [2.0-2.5]    & 2.2 [2.1-2.4]        \\
kT$_{1}$ (keV)                        & 2.7 [2.4-2.9]    & 0.8 [0.7-0.8]       \\ 
kT$_{2}$ (keV)                        & ...              & 3.6 [3.2-4.1]      \\
$\tau$ (10$^{11}$ s cm$^{-3}$)        & 1.4 [1.1 - 1.8]  & ...                \\
EM$_{1}$ (10$^{54}$  cm$^{-3}$)       & 1.52             & 3.35       \\
EM$_{2}$ (10$^{54}$  cm$^{-3}$)       & ...              & 0.74      \\
Abundances$^a$                        & varied           &  varied             \\
Ne                                    & 4.1 [0.0 - 16.]  & $\leq$5.4 \\
Mg                                    & 1.8 [0.7 - 3.6]  & 2.1 [0.9 - 3.5] \\
Si                                    & 4.1 [3.2 - 5.1]  & 5.3 [4.3 - 6.6]  \\
S                                     & 5.9 [5.0 - 7.0]  & 12. [9.5 - 15.] \\
Ar                                    & 7.1 [4.5 - 9.8]  & 18. [12. - 26.] \\
Ca                                    & 6.4 [1.5 - 11.]  & 17. [7.5 - 27.] \\
Fe                                    & 5.9 [4.3 - 8.2]  & 5.3 [4.2 - 6.5] \\
$\chi^2$/dof                          & 384/330          & 389/328            \\
$\chi^2_{red}$                        & 1.16             & 1.19     \\
F$_{x}$ (10$^{-12}$ ergs cm$^{-2}$ s$^{-1}$)  & 1.50 (14.4)      & 1.52 (11.2)    \\
log L$_{x}$ (ergs s$^{-1}$)           & 32.83            & 32.73     \\
log [L$_{x}$/L$_{wr}$]                & $-$6.4           & $-$6.5    \\
\hline
\end{tabular}

{\em Notes}: Based on simultaneous fits of the pn and combined MOS1$+$2 spectra 
binned to a minimum of 20 counts per bin using XSPEC v. 12.3.0. Tabulated
quantities are the  neutral hydrogen absorption column density (N$_{H}$),
plasma energy (kT), upper limit on the shock ionization timescale 
($\tau$ = n$_{e}t$ where n$_{e}$ is the postshock electron density and
$t$ is the time since the plasma was shocked),
emission measure  (EM), and  element abundances. 
Brackets enclose 90\% confidence ranges. The X-ray flux F$_{x}$ in the 
0.5 - 10 keV range is the absorbed value followed in parentheses by
the unabsorbed value. The  X-ray luminosity L$_{x}$ is the unabsorbed
value in the 0.5 - 10 keV range.  A distance of 630 pc is assumed. 
The ratio L$_{x}$/L$_{wr}$ assumes a luminosity for the WR star
log [L$_{\rm wr}$/L$_{\odot}$] = 5.65 (Morris et al. 2000).  

$^a$Abundances are relative to the solar values of 
Anders \& Grevesse (1989) as set in XSPEC using the 
command {\em abund angr}. A value of unity means solar abundance.
The abundances of He, C, N, O were held fixed 
at WN values (see text): He = 25.6, C = 0.9, N = 140., O = 0.9.
The adopted He abundance by number relative to hydrogen is 
25.6 $\times$ 9.77e-02 = 2.5 (M00) where 9.77e-02 is the solar
He abundance by number relative to hydrogen.
The Ni abundance was fixed at its solar 
value in the absence of specific information, but fits are 
insensitive to Ni. Spectral lines from Ne, Ar, and Ca are 
weak or absent in the EPIC spectra and their abundances
are thus not well-constrained.

\end{minipage}
\end{table*}

\clearpage
\newpage




\clearpage
\newpage

\begin{figure*}
\begin{center}
\includegraphics[width=95mm,angle=270.]{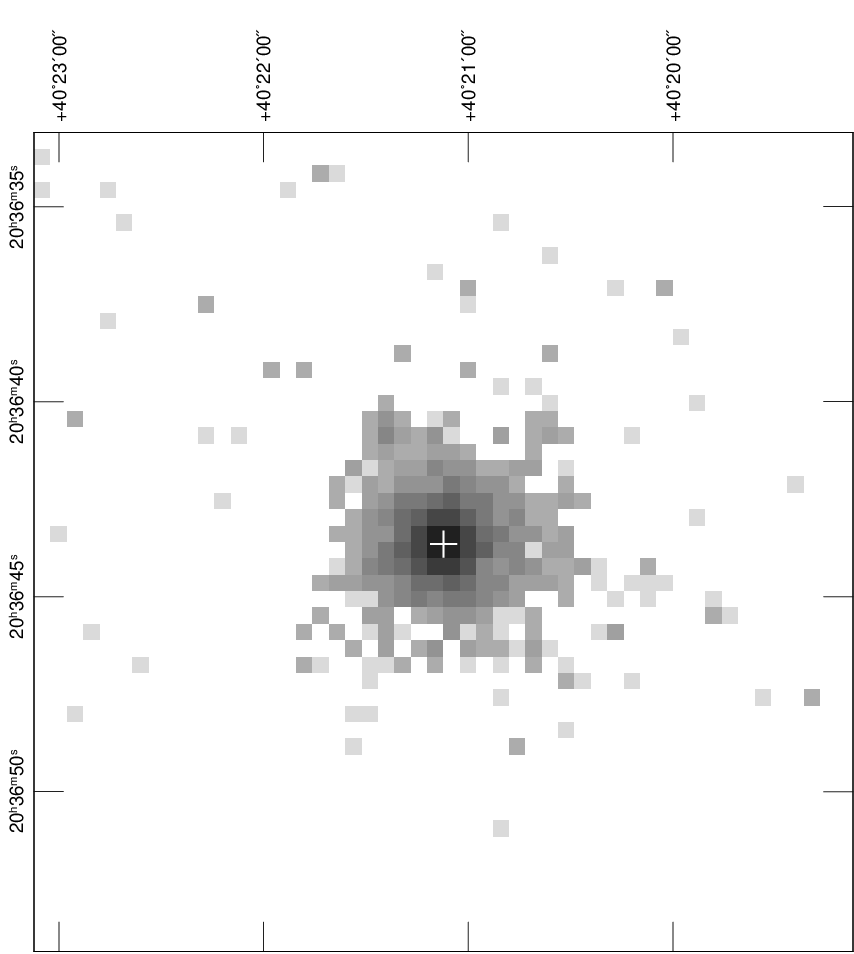}
\vspace*{0.5cm}
\includegraphics[width=95mm,angle=270.]{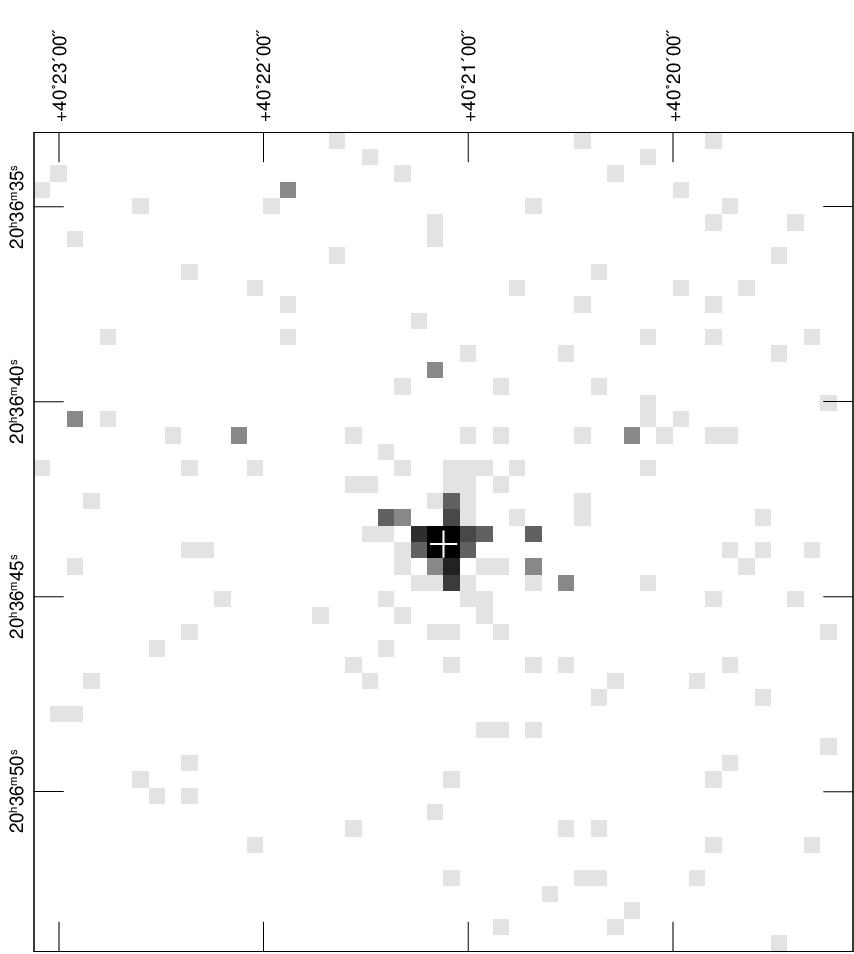}
\end{center}
\caption{{\em Top}: EPIC pn  image of the region near WR 147 in 
                    the 0.5 - 7 keV energy range on a log intensity
                    scale, rebinned to a pixel size of 4.$''$8 for display. 
                    Coordinates are J2000.0 and the cross marks the
                    Simbad position of WR 147.
         {\em Bottom}: Same as above only restricted to the
                       6.3 - 7 keV range, dominated by Fe K$\alpha$
                       emission.} 
\end{figure*}

\clearpage
\newpage

\begin{figure*}
\includegraphics[width=80mm,angle=270.]{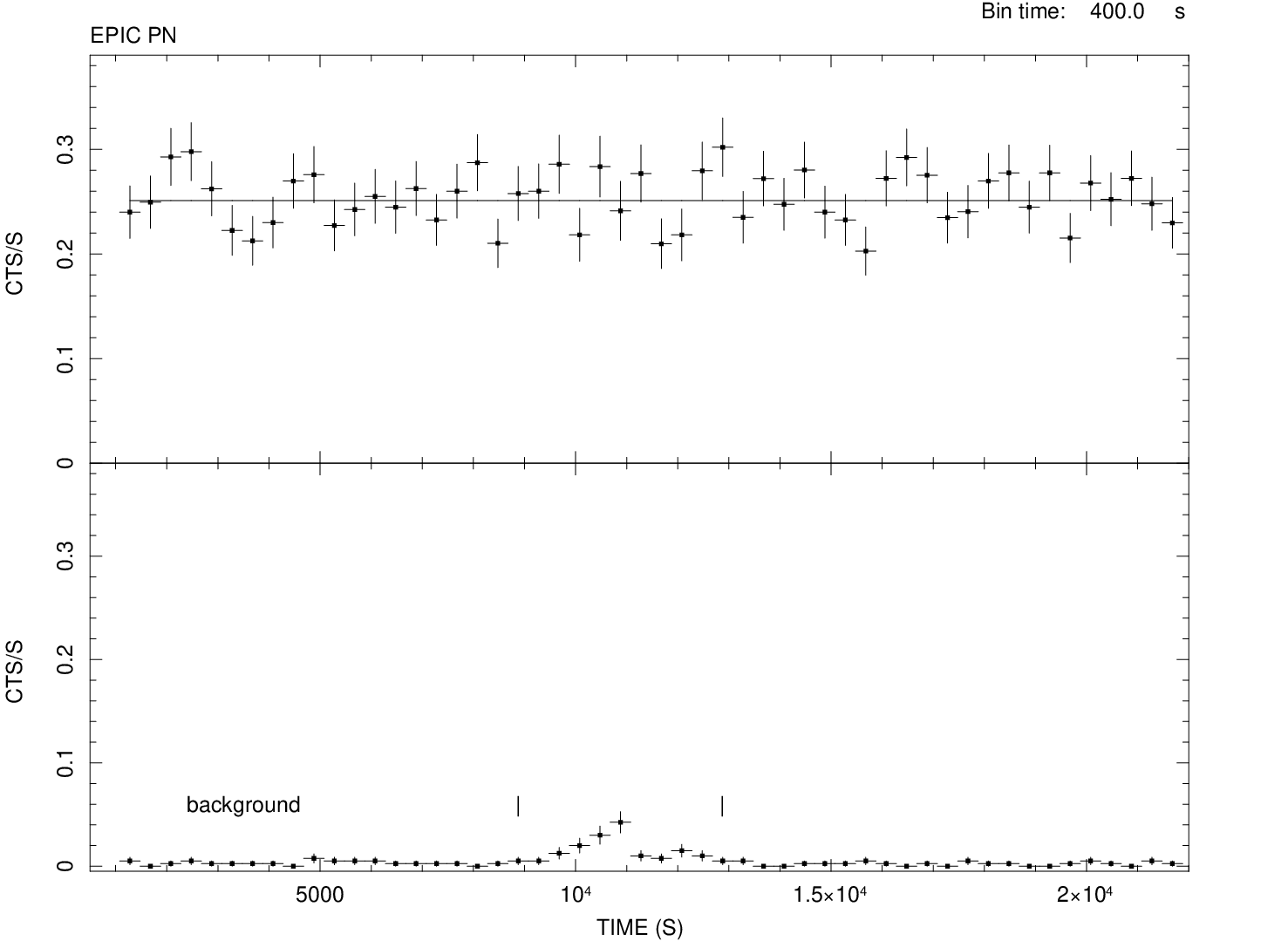}
\caption{Background-subtracted EPIC pn light curve of
         WR 147 extracted from a circular region of
         radius 25$''$ centered on WR 147 and binned
         at 400 s intervals. (top panel). The energy 
         range is  0.5 - 7 keV and the solid line is
         a constant count-rate fit to the data. Error
         bars are $\pm$1$\sigma$.
         The bottom panel shows the background 
         light curve extracted near the source in
         the 0.5 - 7 keV range. Vertical bars
         mark the high-background interval that 
         was excluded from spectral analysis.}
\end{figure*}

\clearpage

\begin{figure*}
\includegraphics[width=80mm,angle=270.]{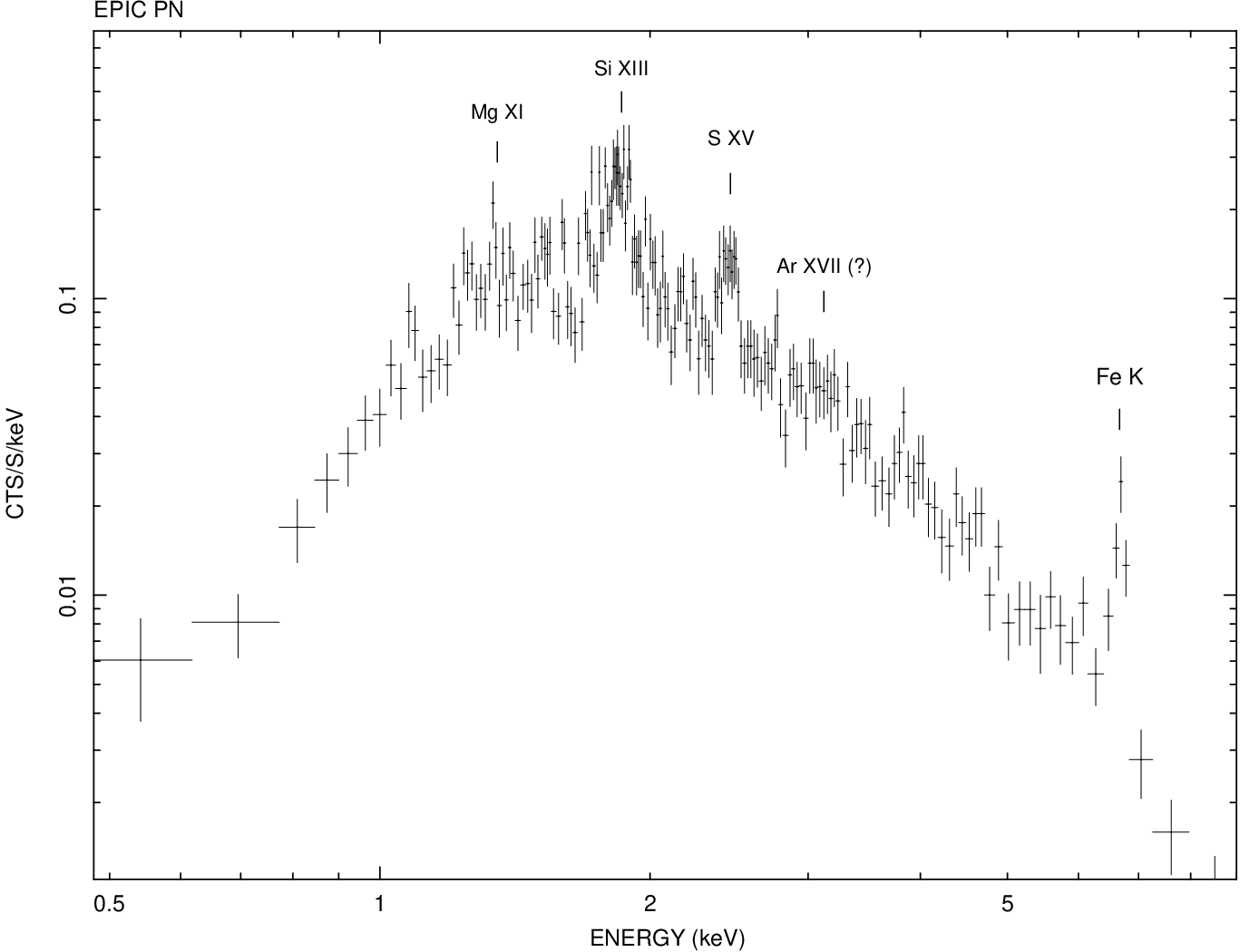}
\caption{Background-subtracted time-filtered EPIC 
         pn spectrum of WR 147 (4436 net counts)
         extracted from a circular region of
         radius 25$''$ and binned to a minimum of
         20 counts per bin.}
\end{figure*}

\begin{figure*}
\includegraphics[width=80mm,angle=270.]{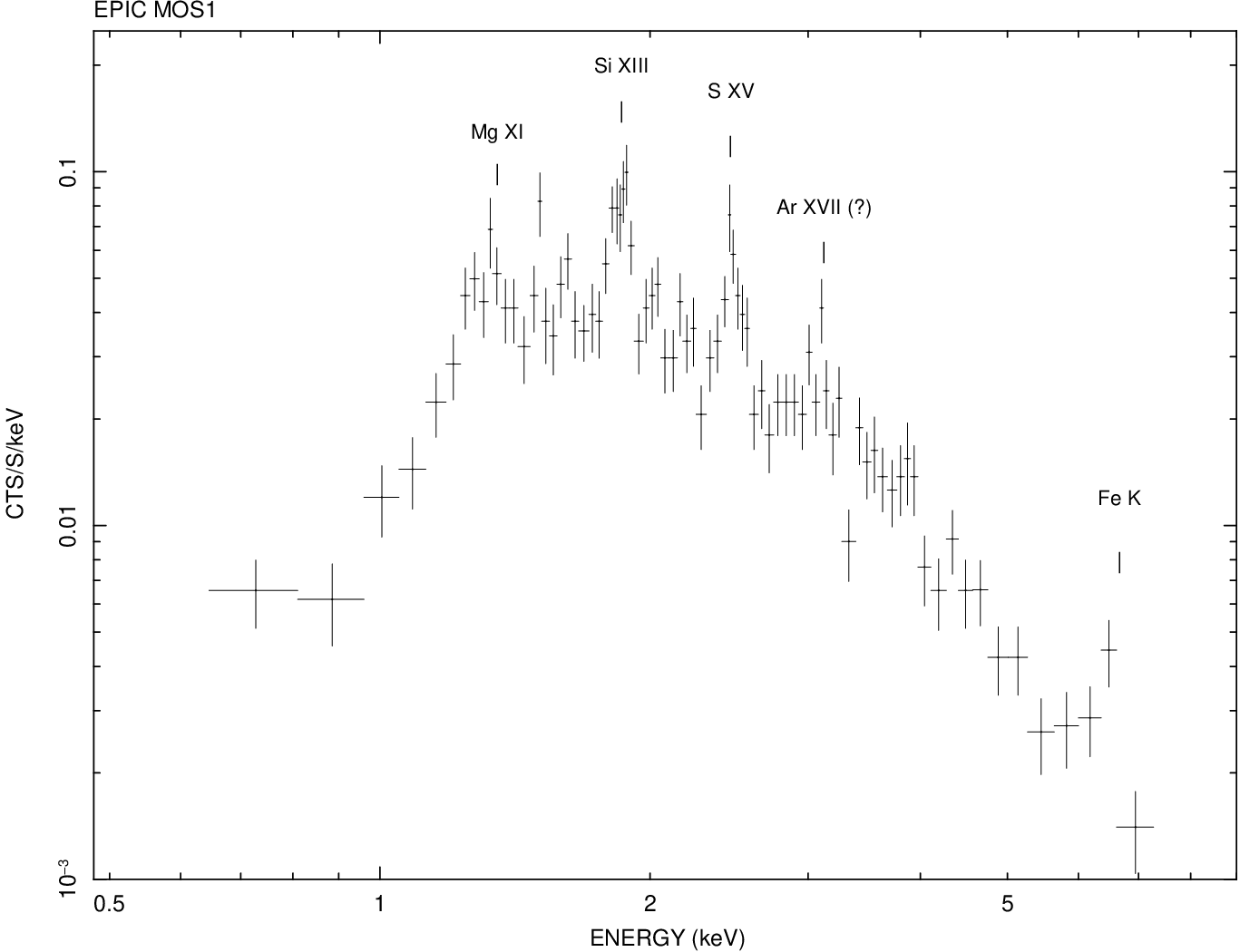}
\caption{Background-subtracted time-filtered EPIC MOS1 spectrum of
         WR 147 (2085 net counts) extracted from a circular region of
         radius 25$''$ and binned to a minimum of
         20 counts per bin.}
\end{figure*}

\clearpage

\begin{figure*}
\includegraphics[width=80mm,angle=270.]{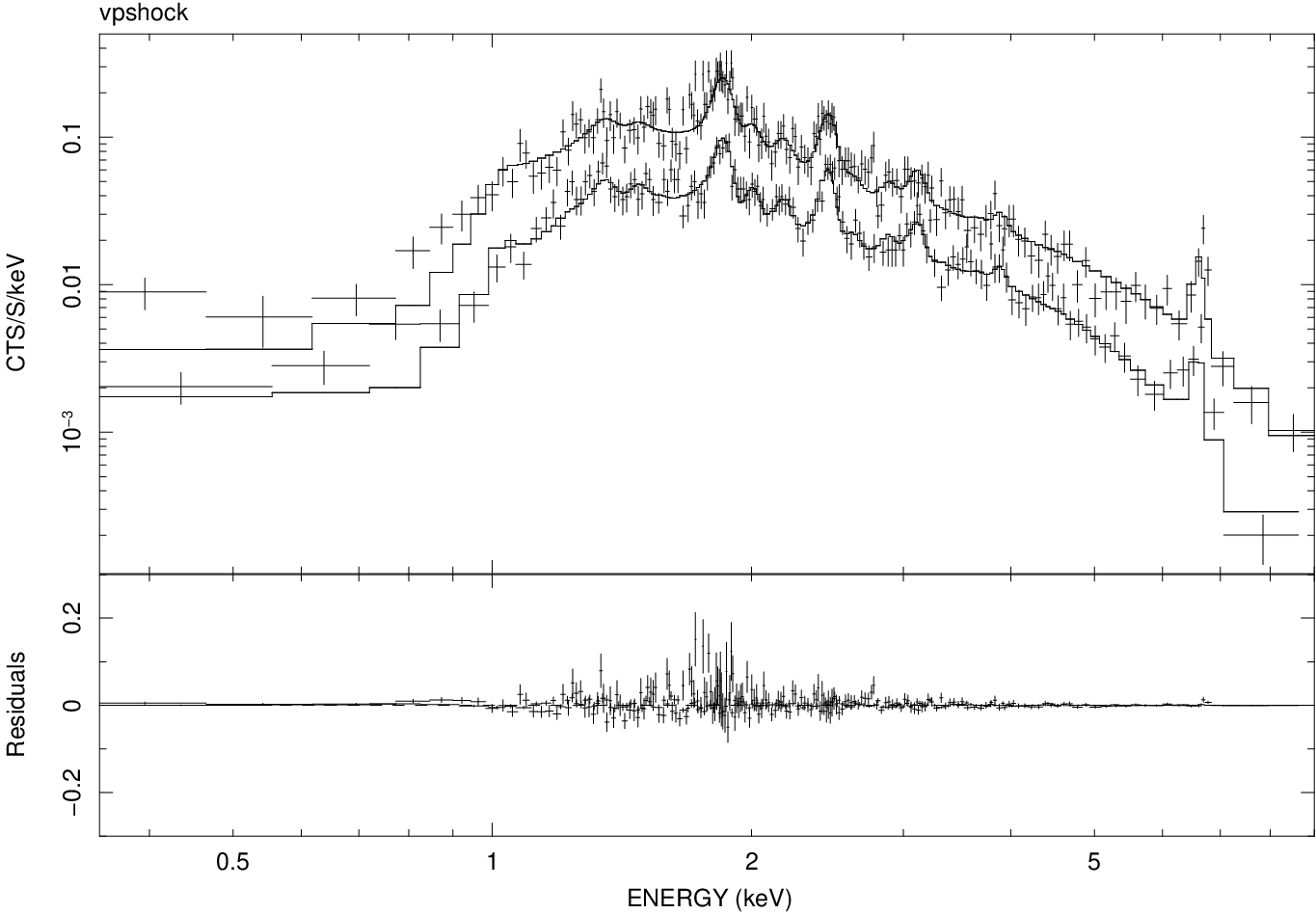}
\caption{Best-fit plane-parallel shock  model $vpshock$ (solid line)
         overlaid on the pn (top) 
         and MOS1$+$2 spectra (bottom). Fit parameters are given
         in Table 1. Bottom panel shows fit residuals. Error bars
         are 1$\sigma$.}
\end{figure*}

\begin{figure*}
\includegraphics[width=80mm,angle=270.]{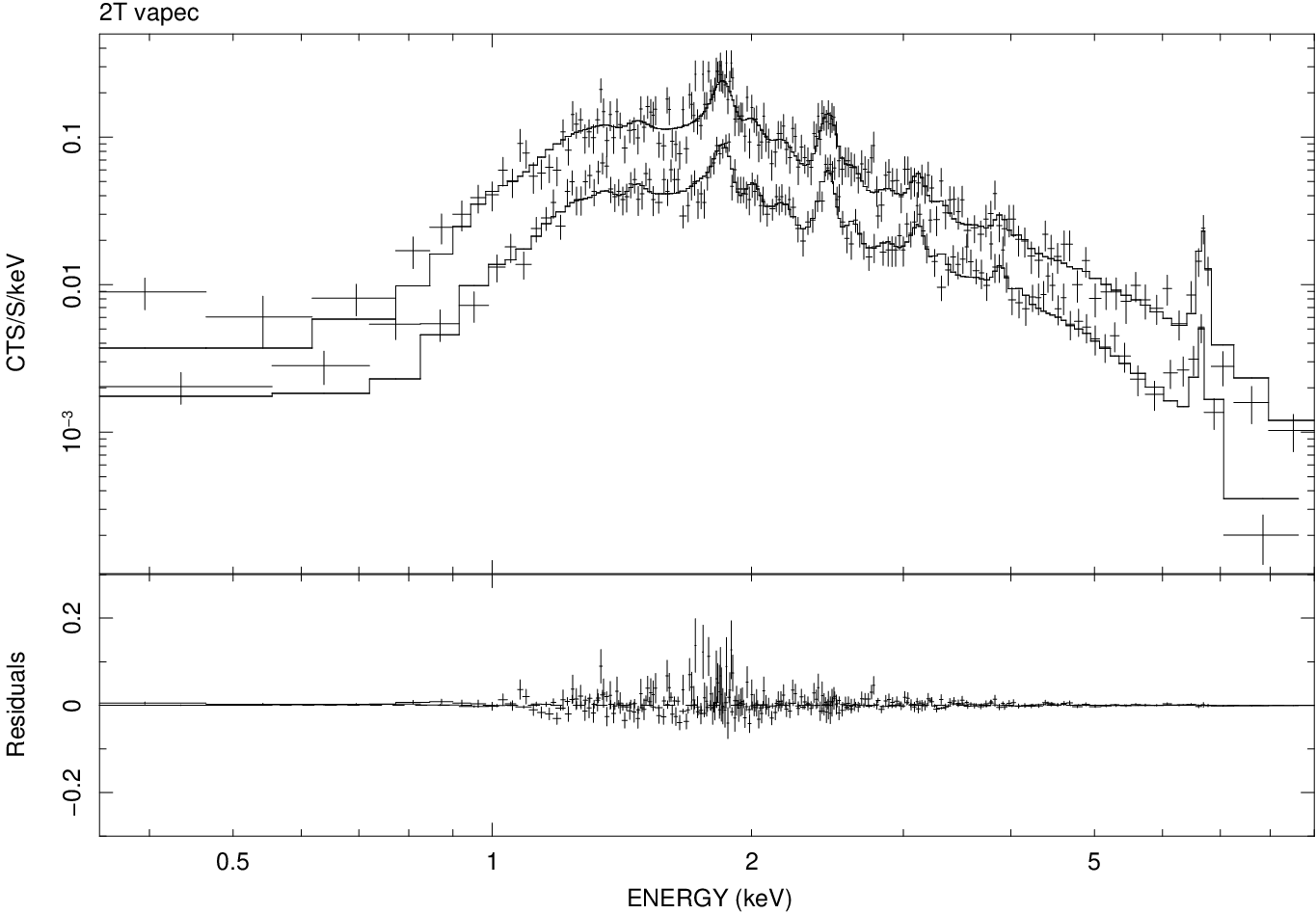}
\caption{Best-fit two-temperature optically thin plasma  model 2T $vapec$
         (solid line) overlaid on the pn (top) 
         and MOS1$+$2 spectra (bottom). Fit parameters are given
         in Table 1. Bottom panel shows fit residuals. Error bars
         are 1$\sigma$.}
\end{figure*}

\clearpage

\begin{figure*}
\includegraphics[width=80mm,angle=270.]{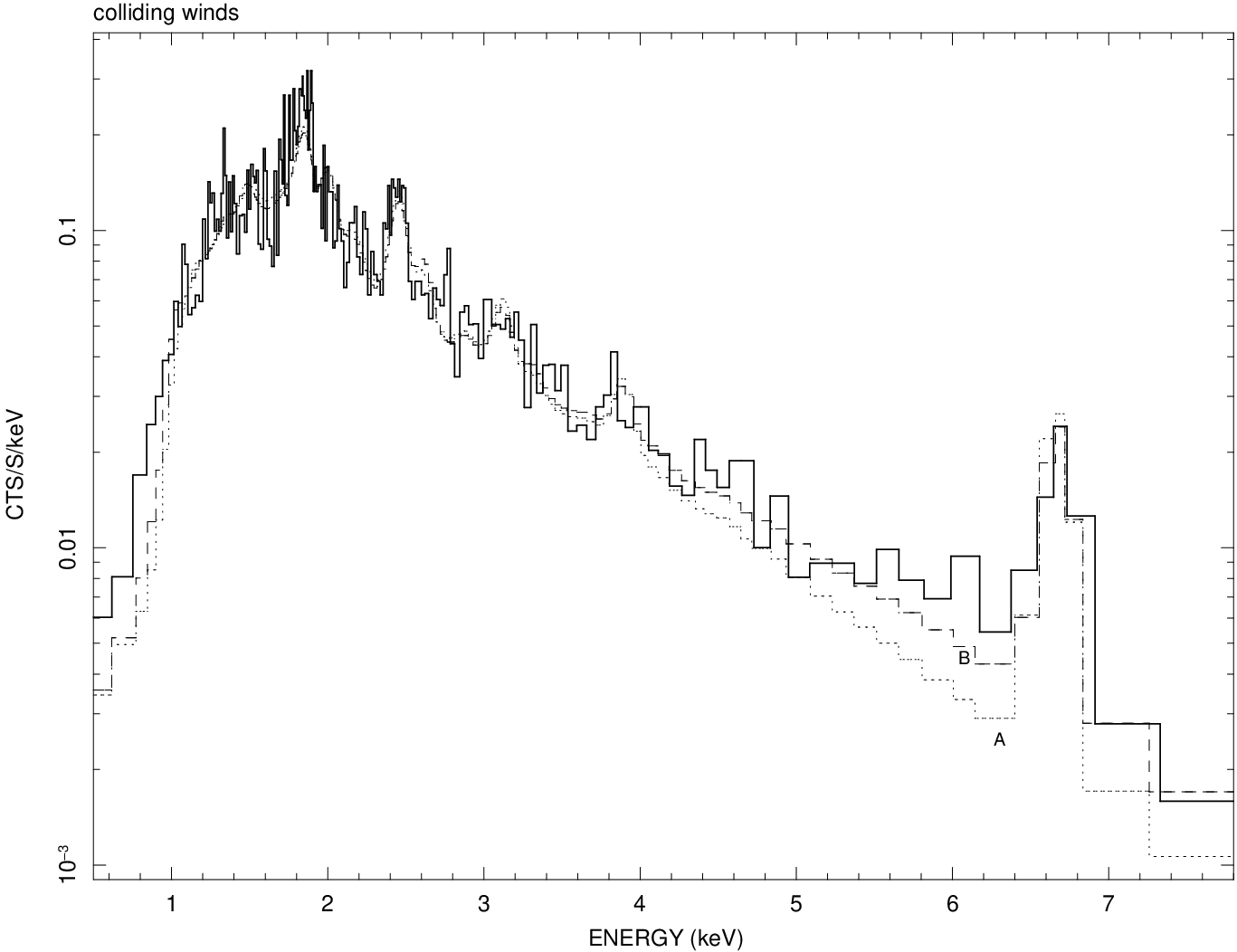}
\caption{Colliding wind shock models overlaid on the pn spectrum of 
         WR 147 (solid line).
         The dotted line (A) uses the nominal  mass-loss parameters
         given in Section 4.2 and a wind momentum ratio $\eta$ = 0.028.
         The dashed line (B) assumes 30\% higher wind velocities with
         values $v_{\infty}$(WR) = 1235 km s$^{-1}$ and
         $v_{\infty}$(OB) = 2080 km s$^{-1}$ and leaves the wind 
         momentum ratio unchanged.  Error bars have
         been omitted for clarity and the X-axis scale is linear to
         accentuate the high energy portion of the spectrum.}
\end{figure*}

\clearpage

\newpage
\bsp
\label{lastpage} 
\end{document}